\documentclass[journal]{IEEEtran}
\usepackage[usenames]{color}
\usepackage{epsfig}
\usepackage{graphics}
\usepackage{amsmath}
\usepackage{amssymb}
\usepackage{multirow}
\usepackage{cite}
\usepackage{array}
\usepackage{pslatex} 
\usepackage{url}
\usepackage{caption}
\usepackage{lineno}
\usepackage{graphicx}  
\usepackage{setspace}
\usepackage{tikz}
\usepackage{hhline}
\usepackage{mathtools}
\usepackage[letterpaper]{geometry}
\ifCLASSINFOpdf
\else
\fi
\usepackage[tight,footnotesize]{subfigure}

\usepackage[english]{babel}
\usepackage[utf8]{inputenc}
\usepackage{fancyhdr}
\begin{document}
\title{\textcolor{purple}{Affective Brain-Computer Interfaces: A Tutorial to Choose Performance Measuring Metric}}

\author{Md Rakibul Mowla*, Rachael I~Cano, Katie J~Dhuyvetter, and David E. Thompson 
\thanks{This material is based upon work supported in part by Kansas State University faculty startup funds and in part by the National Science Foundation under Award No. 1910526.

	\par Md Rakibul Mowla and David E. Thompson are with the Department of Electrical \& Computer Engineering, Kansas State University, Manhattan, KS 66506 USA. K. J. Dhuyvetter was with the Department of Electrical and Computer Engineering, Kansas State University, Manhattan, KS, 66506, USA. R. I. Cano is with the Department of Mathematics, Kansas State University, Manhattan, KS, 66506.  (*correspondence e-mail: rakibulmowla@ksu.edu).
}}



\maketitle\thispagestyle{fancy}

\begin{abstract}
Affective brain-computer interfaces are a relatively new area of research in affective computing. Estimation of affective states can improve human-computer interaction as well as improve the care of people with severe disabilities. To assess the effectiveness of EEG recordings in recognizing affective state, here we used data collected in our lab as well as the publicly available DEAP database. We also reviewed the articles that used the DEAP database and found that a significant number of articles did not consider the presence of the class imbalance in the DEAP. Failing to consider class imbalance creates misleading results. Further, ignoring class imbalance makes the comparing results between studies impossible, since different datasets will have different class imbalances. Class imbalance also shifts the chance level, hence it is vital to consider class bias while determining if the results are above chance. To properly account the effect of class imbalance, we suggest the use of balanced accuracy as a performance metric and its posterior distribution for computing credible intervals. For classification, we used features mentioned in the literature and additionally theta beta-1 ratio. Results from DEAP and our data suggest that the beta band power, theta band power, and theta beta-1 ratio are better feature sets for classifying valence, arousal, and dominance, respectively.  
\end{abstract}

\begin{IEEEkeywords}
Affective brain-computer interfaces (aBCIs), balanced accuracy, DEAP database, electroencephalogram (EEG), support vector machines (SVMs), emotion classification, performance measurement.
\end{IEEEkeywords}

%
\IEEEpeerreviewmaketitle


\section{INTRODUCTION}\label{sec:intro}
\IEEEPARstart{T}{he} term "affective" is a psychological concept referring to the experience of human emotion or feeling. Brain-computer interfaces (BCIs) are usually defined as a direct means of communication between the brain and external devices or systems which enable the brain signal to control some external activity \cite{wolpaw2000brain}. Yet BCIs also allow investigation of brain activity and analysis of brain state. Affective Brain-Computer Interfaces (aBCIs) can be defined as a human affect estimation system from brain signals using BCIs. The interest in automatic detection of people's affective states has increased over the last few decades. Studies have shown that affective states play an important role in human decision making \cite{forgas1995mood}. The ability to manage one's affective states is also related to the ability of logical reasoning, learning and extracting important information \cite{salovey1990emotional}. According to Goleman's model of emotional intelligence, having knowledge of your own affective states is a key factor behind personal and professional success \cite{goleman1996emotional}.

However, estimation of the affective state is a difficult task for several reasons.  Human subjects do not always reveal their true emotions, and often inflate their degree of happiness or satisfaction in self-reports \cite{strack1989salience}. Additionally, there is some ambiguity in understanding and defining affective states \cite{picard2001toward}. 

Facial expression analysis is one of the most popular methods \cite{pantic2000automatic} for estimating affective states, but it is possible to deliberately fake facial expressions unrelated to one's true inner affective state. Therefore, as Picard argued, the estimation may have a high error rate if someone has the ability to disguise his or her emotion \cite{picard2001toward}. 

\begin{figure}[t!]
     \centering 
    \includegraphics[width=.50\textwidth]{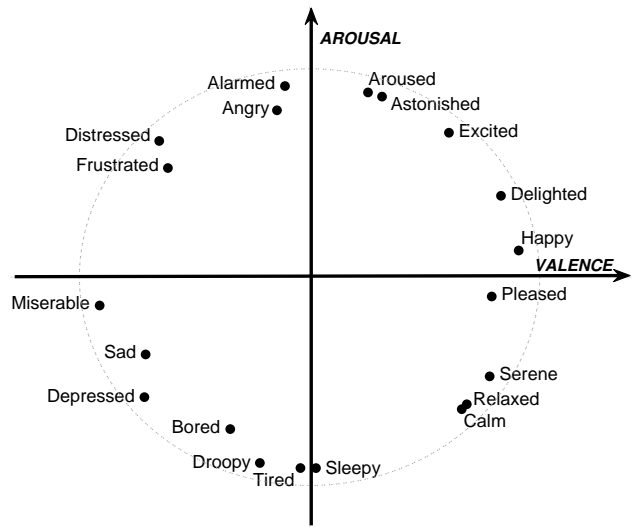}
\caption{An example of the circumplex model where emotions are expressed in the valence and arousal dimensions. Valence refers to how pleasant or unpleasant an emotion is, and arousal refers to how exciting or boring it is. Words are placed according to direct circular scaling coordinates for 28 affect words from Russel's article \cite{russell1980circumplex}.} \label{fig:1}
\end{figure}

With the improvements in brain imaging techniques, there is a growing interest in relationships between affective states and brain activities. Investigating affective states using electroencephalogram (EEG) is becoming popular among researchers because EEG is one of the most convenient, non-invasive forms of recording brain activity. EEG also has high temporal resolution, which makes it a preferable candidate for fast affective state estimation \cite{niemic2002studies}. Before using EEG-based BCIs to estimate affective states, one major challenge is to model affective states in a measurable and understandable scale. A current, widely accepted affective state model is the circumplex model of affect (Figure \ref{fig:1}), which was initially proposed by J. A. Russel \cite{russell1980circumplex}. Finding distinct physiological patterns for each affective state has also always been a major topic of interest for affective computing researchers \cite{cacioppo1990inferring}. Picard argued that emotion consists of more complex, underlying processes rather than outward physiological expression \cite{picard2001toward}. 

Interest in EEG-based emotion recognition has increased over time and is still growing. Searching "EEG emotion recognition`` in Google scholar gives 115000 results in March 2020. Among them, there are 2100 just in the first quarter of 2020. Because these projects rely on individuals' emotional responses, the distribution of affective states (classes) is often uneven. However, most of these articles do not mention the class imbalance percentage but instead only report classification accuracy as a performance measuring metric. This creates a serious ambiguity and makes the results incomparable between works. For example, a publicly available database for emotion recognition known as the DEAP database \cite{koelstra2012deap} has been cited over 1600 times on March 2020, and using "EEG emotion recognition`` search keywords within the DEAP-citing articles gives more than 1330 results. Out of those 1330 articles, at least 170 articles have included the DEAP dataset in their analysis. Out of those 170 articles, only approximately thirty-three articles mentioned or considered class imbalance. Classification accuracy, without considering class imbalance, is misleading for reasons we will present in this paper. Additionally, out of those 170 articles, only approximately 30 articles discussed statistical significance. This raised a few serious research questions: 
\begin{enumerate}
    \item Are those classification accuracies better than unskilled classifiers?
    \item If so, are those accuracies significantly better than chance? 
    \item In the presence of class imbalance, what is the correct chance level?  
    \item What performance evaluation metric should be used in affect classification?
\end{enumerate}
The main goal of this work is to investigate these questions.  As a case study, we will use our investigations into EEG-based detection of binary (high/low) valence, arousal, and dominance in response to different sets of stimuli. For this investigation, we use both our own data as well as the previously mentioned, publicly available DEAP database \cite{koelstra2012deap}. 

Affective states can be elicited through visual \cite{lang2008iaps}, auditory \cite{lang1999iads}, and audio-visual stimuli\cite{baveye2015liris}, among other methods. The emotional experience is more profound when visual presentations are combined with auditory stimuli, intermediate under visual stimuli and minimal during auditory stimuli \cite{guntekin2014review}. In our experiment, we used visual stimuli, the International Affective Picture System (IAPS) \cite{lang2008iaps}, to evoke emotions. The DEAP database used audio-visual stimuli.

\section{Related Work}
In the field of affect recognition, a huge number of studies have been conducted on emotion recognition using EEG signals. With the improvement of dry electrodes, EEG is nearing or at the point of being a practical, out of the lab solution for affect recognition. More detailed EEG-based emotion recognition reviews can be found in \cite{wagh2019electroencephalograph,garcia2019review}. 
One major problem in EEG-based emotion recognition research is the lack of publicly available datasets. Consequently, researchers use their own data and as a result studies become more difficult to compare. To solve this problem, a few researchers developed publicly available datasets including the  DEAP \cite{koelstra2012deap}, USTC-ERVS\cite{wang2014hybrid} and MAHNOB-HCI datasets\cite{soleymani2011multimodal}. Among these datasets, the DEAP is the most cited and used for emotion recognition. Thus, we were motivated to use the DEAP dataset in this work. 

Studies where DEAP was used as the benchmark dataset mostly used support vector machine (SVM) \cite{piho2018mutual,li2018exploring,soleymani2017toolbox,zheng2017identifying,wang2017content,ozerdem2017emotion,verma2017affect} for classification. The second most-used classification technique was the k-nearest neighbor (kNN) classifier \cite{piho2018mutual,zheng2017identifying,ozerdem2017emotion}. Other classification techniques, such as deep convolutional neural network\cite{li2017human}, decision tree\cite{garcia2016application}, linear discriminate analysis (LDA)\cite{al2018anytime}, logistic regression\cite{zheng2017identifying}, discriminative graph regularized extreme learning machine (GELM)\cite{zheng2017identifying}, back-propagation neural networks (BPNN)\cite{purnamasari2017development}, probabilistic neural networks (PNN)\cite{purnamasari2017development}, and multilayer perceptron (MLP)\cite{verma2017affect} have also been used to classify emotion on the DEAP dataset.  Features used in these studies are statistical features: mean, standard deviation, variance, zero crossing rate\cite{liu2018emotion},\cite{liu2018emotion,torres2017svm,verma2017affect,menezes2017towards}, Hjorth parameters\cite{li2018exploring,mert2018emotion}, fractal dimension\cite{liu2018emotion,nakisa2018evolutionary}, Shannon entropy\cite{liu2018emotion}, spectral entropy\cite{liu2018emotion,verma2017affect}, kurtosis\cite{hemanth2018brain}, skewness \cite{yin2017cross}, different EEG band powers\cite{torres2017svm,yoon2013eeg}, relative power spectral density (PSD) for delta, theta, alpha, beta and gamma frequency bands \cite{wang2015emotion}, differential entropy (DE), differential asymmetry (DASM), rational asymmetry (RASM), asymmetry (ASM)\cite{zheng2017identifying}, wavelet coefficients\cite{ozerdem2017emotion}, and higher order crossings (HOC)\cite{piho2018mutual}. 

In the DEAP dataset, emotions are expressed in valence, arousal, and dominance dimensions on discrete 9-point scales. To design the classification model those scales need to be labeled. Here also, inconsistencies exist between different studies. Not only are different numbers of classes chosen by different groups, but even within the same number of classes the thresholds are different.  In these previously mentioned studies on the DEAP, classification labels were created by splitting the ratings into  3-class (1-3:negative, 4-6:neutral, and 7-9:positive) \cite{jirayucharoensak2014eeg}, 3-class (1-4.5:negative, 4.5-5.5:neutral, 5.5-9:positive)\cite{verma2017affect}, 2-class (High/low, 4.5-9: high)\cite{daimi2014classification}, 2-class (negative $\leq 5 <$ positive)\cite{wang2015emotion}, 2-class (negative $< 5 \geq$ positive)\cite{padilla2016emotion,gupta2016relevance,wang2017content}, and 2-class (1-3: low and 7-9: high)\cite{menezes2017towards}. Hence, the class imbalance in all these studies are different based on their individual approach when generating class labels. 

Even though all these above-mentioned studies used the DEAP dataset, where significant class imbalance exists, very few studies have considered it while reporting results. Studies where class imbalance was considered mainly reported the F1 score\cite{koelstra2012deap,yin2017recognition,soleymani2017toolbox,garcia2016application,padilla2016emotion} and a few other studies used receiver operating characteristic (ROC)\cite{menezes2017towards,piho2018mutual}, area under ROC (AUC) \cite{li2018exploring} and balanced accuracy\cite{clerico2018eeg} along with accuracy metric. But AUC can be a misleading metric for a comparative study especially in the presence of variable class imbalance \cite{lobo2008auc} and computing the F1 score for multiclass classification is also not straightforward. For multiclass problems, F1 can be computed using macro-averaging or micro-averaging \cite{van2013macro}. The difference between macro- and micro-averaged F1 can be large; if studies do not report which was used then comparing results is impossible.  For example, \cite{gupta2016relevance} reported classification accuracies of 67\% and 69\% and F1 scores of 0.67 and 0.69 for valence and arousal, respectively. It is not clear how these F1 scores were calculated. F1 scores for both classes were not considered in that study which makes the study incomparable and provides misleading results. 

To eliminate those above-mentioned problems we are suggesting to use balanced accuracy as the classification performance evaluation metric in high/low valence, arousal and dominance classification. To our knowledge, this has only been used in \cite{clerico2018eeg}.  However, that study did not include the computation of credible intervals for balanced accuracy; here in this study we will further discuss using the posterior distribution of balanced accuracy to compute  credible intervals and perform statistical significance testing. 

\section{Data Description}
In this work, we have used data from the publicly available DEAP dataset and EEG recordings from our lab. 
\subsection{Database for emotion analysis using physiological signals (DEAP)}
The DEAP is a publicly available, multimodal dataset consisting of 32-channel EEG, electrooculography (EOG), electromyography (EMG), galvanic skin response, respiration, plethysmograph, and temperature data \cite{koelstra2012deap}. These signals were collected from thirty-two healthy participants, with an equal male-female ratio and an average age of 24.9 years. Data were recorded at a sampling rate of 512Hz and then pre-processed. 

Minute-long music videos were used as emotional stimuli. After each video, participants were provided enough time to rate those videos for valence, arousal, and dominance on a discrete 9-point scale using self-assessment manikins (SAM) \cite{bradley1994measuring}. Each participant viewed forty videos.

\subsection{Data collected at Brain and Body Sensing (BBS) lab}
\begin{figure}[ht!]
     \centering 
    \includegraphics[width=.50\textwidth]{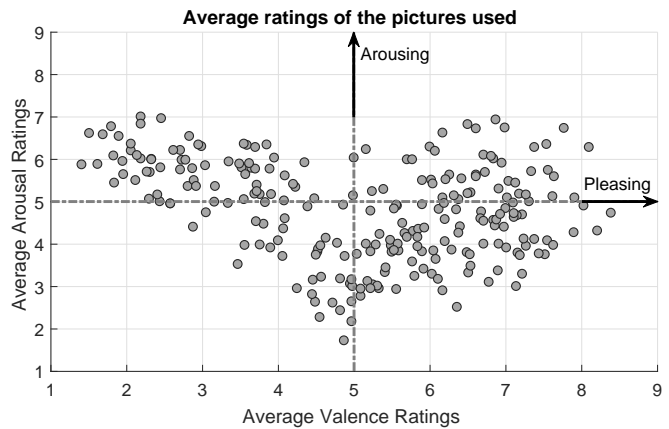}
\caption{Visualization of average valence and arousal ratings (from the IAPS manual) \cite{lang2008iaps} of picture sets used to collect data at the BBS lab.} \label{fig:2}
\end{figure}

The BCI2000 \cite{schalk2004bci2000} system was used to present picture stimuli to the participants. Each picture was displayed for 6.7 seconds, followed by a 20.8s pause for participants' self-report. A total of 244 pictures were selected from IAPS \cite{lang2008iaps} images; the average valence and arousal ratings reported in the IAPS manual of the selected pictures are shown in figure \ref{fig:2}. Pictures were presented in six blocks, with breaks for participant comfort. EEG data were recorded using a Cognionics Mobile-72 EEG system with a sampling frequency of 600Hz. The Mobile-72 EEG system is a high-density mobile EEG system with active Ag/AgCl electrodes placed according to the modified 10-20 system. Reference and ground were on the right and left mastoids, respectively. 
In total, we had nine participants. Data from two participants have been excluded due to one data entry error and one battery failure. All participants were healthy college students with an age range of 21 to 22 years. Each participant was shown 244 pictures through two or three different sessions. Each participant rated each stimulus for valence, arousal, and dominance on a discrete 5-point scale using self-assessment manikins (SAM) \cite{bradley1994measuring}.

\subsection{Pre-processing}
For the DEAP, both raw and pre-processed data are available for use. In this work, we will use this \textsc{Matlab}-ready preprocessed version of the data. Pre-processing includes common-average referencing, down-sampling to 128Hz, band-pass filtering with the cut-off frequency at $(4.0-45.0)$ Hz, and eye blink artifact removal via independent component analysis. The data contain 32 channels of EEG plus an additional eight channels of other physiological signals and the length of the time segment for each trial is 60 seconds. We have only used EEG recordings for classification.  Data were then transformed into scalp surface Laplacian or current-source density (CSD) because it has been argued that CSD transformation gives a more sensitive index of individual variations in frontal asymmetry than other EEG recording montages and also helps to reduce non-frontal contributions to the frontal asymmetry \cite{velo2012should,allen2015frontal}.

The data collected at the BBS lab was filtered using a finite impulse response (FIR) bandpass filter at $(4.0-45.0)$ Hz. Data were then transformed into scalp surface Laplacian or current-source density (CSD). To transform the EEG recordings into surface Laplacian, we used the CSD toolbox \cite{kayser2006principal} which provides a \textsc{Matlab} implementation and uses the spherical spline algorithm \cite{perrin1989spherical} to estimate the surface Laplacian.

\section{Methods}\label{meth}
In our study, we will use $x(t) \in \mathcal{R}^T$ as the time series of a recording from a single electrode with $N$ samples. The first and second derivatives of $x(t)$ with respect to time are $x'(t)$ and $x''(t)$, respectively. Standard deviation of $x(t)$, $x'(t)$ and $x''(t)$ are denoted as $\sigma_{x}$, $\sigma_{d}$, and $\sigma_{dd}$, respectively. Class labels are denoted by $c \in \{1,2,\ldots,C\}$ and predicted class labels are denoted by $y$ when classifying. $\mathbb{H}$ denotes entropy.  

\subsection{Feature sets}
\subsubsection{Frequency domain features}
Power spectral density and signal power at different frequency bands are popular for EEG-based affective state classification and have been used as features in several studies (e.g.  \cite{lin2010eeg,jenke2014feature}). Spectral density and band powers can be computed using various algorithms, including Fast Fourier Transform, short-Time Fourier Transform, or Welch's power spectral density estimations algorithm. Here, we have used Welch's power spectral density (PSD) estimation method \cite{welch1967use} and then computed power in each band powers from the resulting PSD. The frequency ranges used for EEG bands varies slightly between different studies. In our analysis, the frequency ranges we have used are theta: (4-8) Hz, alpha: (8-12) Hz, Beta-1: (12-18) Hz, Beta-2: (18-30) Hz, Gamma: (31- 63) Hz. 
    
In a few studies, it has been argued that frontal EEG asymmetry can be a moderator and mediator of affective states \cite{coan2004frontal,allen2004issues}. 
Frontal alpha asymmetry is mostly used as a discriminator between depressed and healthy individuals \cite{van2017frontal}. However, it also can be used for affective state classification. Here, we will use both frontal EEG asymmetry (1-50 Hz) and frontal alpha  asymmetry (8-12 Hz) as features for classifying affective states. If $R_p$ represents the signal power of electrodes located at the right frontal lobe and $L_p$ represents the signal power of electrodes located at the left frontal lobe then frontal EEG asymmetry can be calculated from 
\begin{equation}\label{faa1}
    \text{Frontal asymmetry} = \ln \big(\frac{R_p}{L_p}\big)
\end{equation}
Another form of the frontal asymmetry is the normalized version of equation (\ref{faa1}) and is written as 
\begin{equation}\label{faa2}
    \text{Frontal asymmetry} = \ln \big(\frac{R_p-L_p}{R_p+L_p}\big)
\end{equation}
Here, we have used equation (\ref{faa1}) to find the frontal asymmetry. We have computed separately the frontal asymmetry index (FAI) and frontal alpha asymmetry index (FAAI). The frequency range of $0-64$Hz is used to compute FAI and the alpha band is used for FAAI. 

We also used frontal theta beta ratios (TBR) as frequency domain features even though TBR has not been used previously for affective classification. But it has been reported to be related with affective traits \cite{putman2010eeg}. To compute the frontal TBR we used equation (\ref{tbr})
\begin{equation}\label{tbr}
    \text{TBR} = \ln \big(\frac{\theta_p}{\beta_p}\big)
\end{equation}
here $\theta_p$ represents the theta band power and $\beta_p$ represents the beta band power of electrodes located at the frontal lobe. Frequency ranges for beta-1 and beta-2 are used in $\beta_p$ to compute TBR1 and TBR2, respectively.

\subsubsection{Hjorth parameters}
Hjorth parameters are time-domain features of EEG recording, proposed by Bo Hjorth \cite{hjorth1970eeg}. Hjorth parameters have been recently used in several studies \cite{jenke2014feature,mert2018emotion} as features for affective state estimation. The parameters are Activity, Mobility, and Complexity. Activity is simply the variance of the time signal. If the signal is denoted as $x(t)$, then Activity $=\sigma_{x}^2$ and is the measure of the squared standard deviation of amplitudes. Mobility measures the standard deviation of the slope with respect to the standard deviation of the amplitude. Mobility is defined as the square root of the ratio between the variances of the first derivative and the time signal. Complexity is a measure of how much the time signal deviates from a pure sine shape and is defined as the ratio between the mobility of the first derivative of the time signal and the mobility of the time signal. 
\begin{align*}
    \text{Mobility} &= \frac{\sigma_{d}}{\sigma_{x}}\\
    \text{Complexity} &= \frac{\sigma_{dd}/\sigma_{d}}{\sigma_{d}/ \sigma_{x}}
\end{align*}
Here, we have used mobility and complexity as features. For each trial, there will be an equal number of mobility and complexity values and the number equals the EEG electrode number.

\subsubsection{Entropy}
Entropy is a measure of disorder in a system. In the case of EEG, entropy measures the irregularity in the signal. Spectral entropy of EEG recordings has been used to discriminate different affective states in other studies \cite{vakkuri2004time} and it recently has been used in recognition of emotional states \cite{zheng2017identifying}. In this work, we will use spectral entropy (SE), which is the normalized Shannon entropy of the power spectrum. 
\begin{equation}\label{entropy}
    \text{Spectral Entropy} = -\frac{\sum_{i=1}^Np(X=i)\log_2 p(X=i)}{\log_2N}
\end{equation}
where $X$ is denoting the power spectrum of the time series $x(t)$, $p(X)$ is the spectral distribution such that $\sum_{i=1}^Np(X=i)=1$, and $N$ is the number of frequency bins. 

 \subsubsection{Feature sets}
 For the valence, arousal and dominance classification we used seventeenth different feature sets which are frontal asymmetry index (FAI), frontal alpha asymmetry index (FAAI), theta beta-1 ratio (TBR1), theta beta-2 ratio (TBR2), theta band power (ThetaP), alpha band power (AlphaP), beta band power (BetaP), gamma band power (GammaP), TBR1 and TBR2 together (TBR-C), theta, alpha, beta and gamma band power all together (TABG), Hjorth parameters (Hjorth), entropy (Entropy), power spectral density (PSD), beta alpha ratio (BARatio), all feature sets mentioned previously together (All), and principal components of all feature sets (All-PCA). In case of All-PCA, we used the principal components which consists the 98\% of total variability. Using these different feature sets for all 32 participants have resulted a $17\times 32$ classification results in each affective dimension for each classifier.

\subsection{Classification}
The ultimate goal for emotion estimation is a many-class classification or continuous-output regression.  However, for this initial investigation, we focused on the easier binary classification problem, following multiple literature examples  \cite{daimi2014classification,wang2015emotion,padilla2016emotion,gupta2016relevance,wang2017content,menezes2017towards}. Thus, we use a two-class classification system for valence, arousal, and dominance. Participants in our experiments rated each axis from 1 to 5, we have labeled $ratings < 3$ as low valence, arousal, and dominance and $ratings \geq 3$ as high valence, arousal, and dominance. One participant never rated arousal less than 3, so for this participant (number 6) we shifted the split point from 3 to 4.  In the DEAP database, participants rated each axis from 1 to 9; we have labeled $ratings < 5$ as low and $ratings \geq 5$ as high following the original work \cite{koelstra2012deap} and some other related studies \cite{mohammadi2017wavelet,liu2018emotion,clerico2018eeg}.
 
In this study, support vector machine (SVM) and K-nearest neighbor ($k$NN) classifiers were used to test the affect recognition from EEG data. For our data, we will use 10-fold cross-validation. In case of DEAP data, we will use ``Leave-One-Out" cross-validation technique. Which means at each step of the cross-validation, one sample was used as the test set and the rest were used as training set. The reason of using ``Leave-One-Out" cross-validation in lieu of ``K-fold" cross-validation is to maintain the congruity with other studies \cite{koelstra2012deap,daimi2014classification,soleymani2017toolbox,clerico2018eeg}. These classifiers are the most commonly used techniques among published reports using the DEAP dataset (e.g. \cite{mohammadi2017wavelet, liu2018emotion,clerico2018eeg,wang2015emotion,verma2017affect,menezes2017towards,piho2018mutual,ozerdem2017emotion}. 

\subsubsection{Support vector machines (SVMs)}
SVM uses a kernel trick and a separating hyperplane to create the support vectors. SVMs can be used for both regression and classification.  In  SVMs, with the observation vector $\mathbf{x}$ the predicted class label can be found using \cite{murphy2012machine}
\begin{align}
     & \hat{f}(\mathbf{x}) = \text{sgn}\Big(\hat{w}_0 +\sum_{i=1}^N\alpha_i k(\mathbf{x}_i,\mathbf{x})\Big)
\end{align}
Where, $\alpha_i = \lambda_iy_i$, $\lambda$ is the $\ell_1$ regularization term and $k(\mathbf{x}_i,\mathbf{x})$ is the kernel function. For Gaussian kernel SVM, the kernel function is defined by 
\begin{align}
    k(\mathbf{x}_i,\mathbf{x}) = \exp \Big( -\frac{1}{2} (\mathbf{x}_i - \mathbf{x})^T \Sigma^{-1} (\mathbf{x}_i - \mathbf{x})\Big)
\end{align} 
Here we have used the \textsc{MATLAB} built-in function \texttt{fitcsvm} for SVM classifier with a Gaussian kernel.

\subsubsection{K-Nearest Neighbours (KNN)}
kNN is a simple classification algorithm where an example is classified based on the plurality vote of its $k$ number of nearest neighbors. The nearest neighbours are chosen by a distance metric. Distance metrics can be City block distance, Chebychev distance, Minkowski distance, Euclidean distance or Mahalanobis distance. Here we have used the built-in \textsc{MATLAB} function \texttt{knnsearch} using Euclidean distance with $k=9$ using Euclidean distance. The kernel and hyperparameters for both classifiers are chosen  empirically using a 15\% test set partition strategy.

\section{Performance Metrics}\label{metric}
The most commonly used classification performance measurement metric is accuracy. Nevertheless, accuracy can be misleading, especially with the presence of class imbalance. In these situations, classifiers can learn from class label proportion rather than the features, a property sometimes known as "unskilled classification."  In biased datasets, the unskilled performance is equal to the class imbalance. Thus, the same reported accuracy should be interpreted differently based on class bias. For example, consider a study reporting 80\%  accuracy in a two-class classification. This may be good performance on a balanced dataset but is at or below unskilled classification levels for biases $\geq 80\%$.

Comparing the performance of a similar classification task with different proportions of class labels is difficult. To make this kind of comparison meaningful, researchers suggest using other performance measuring metrics such as the Kappa statistic or area under ROC curve (AUC) for imbalanced data. But since the multiclass ROC curve analysis is not well developed \cite{lachiche2003improving}, AUC is not recommended for multiclass problems \cite{sokolova2009systematic}. Moreover, the accuracy metric is the most widely used, and the most intuitive solution would be to make the accuracy metric meaningful by scaling down the baseline to be the performance of an unskilled classifier. One way to scale the baseline is to compute the balanced accuracy \cite{velez2007balanced} where the accuracy in each class is considered separately.

\subsection{Balanced Accuracy}
If there are $m$ number of classes, the balanced accuracy \cite{velez2007balanced} is defined as  
\begin{equation}
     \textit{Balanced Accuracy} = \frac{1}{m} \sum_{k=1}^m\frac{C_{kk}}{n_k} 
\end{equation}
Here, $n_k$ is the total number of observations in class $k$ and $C_{kk}$ is the number of correctly classified observations in that same class label.

Since our focus is on two-class classification, here, k=2. If the classifier performs equally well on both classes then the balanced accuracy will be exactly equal to the conventional accuracy \cite{velez2007balanced,brodersen2010balanced}. Since balanced accuracy is the average accuracy of each class, it is unaffected by the class imbalance and is more meaningful than the traditional accuracy metric. Further, it has the convenient property that an unskilled classifier always achieves $1/k$ accuracy regardless of class imbalance.


Although the traditional accuracy metric is a scaled binomial random variable, researchers often use a normal posterior distribution to compute credible intervals. The assumption behind the posterior normal distribution comes from the central limit theorem, where for a sufficiently large number of observations ($ n \geq 30$), a binomial distribution can be approximated using the normal distribution. Nonetheless, this approximation becomes unreliable for small $n$. Particularly in the case of imbalanced data, the number of observations for the minority class can be smaller than the required number for the normal approximation. Therefore, finding chance performance and the credible interval of the misclassification rate for balanced accuracy is not as straightforward as it is in the case of traditional accuracy. For the two-class classification case, it is a combination of two separate distributions. In a multi-class scenario, accuracy in each class will have a separate distribution.

\subsubsection{Credible intervals of Balanced Accuracy}
If the probability of predicting correct classes of a classifier denoted by $\mathcal{A}$ with a prior distribution $p(\mathcal{A})$, then the posterior is expressed as $p(\mathcal{A}|\mathcal{D})$ on observed data $\mathcal{D}$. Lets assume $y=1$ and $y=0$ for correct and incorrect predictions, respectively. Now the classification predictions can be written as $y_1,y_2,\ldots,y_n$ which resembles the results of a Bernoulli experiment. So we can write 
\begin{align}
    p(y_k|\mathcal{A}) & = \text{Bern}(y_k|p(\mathcal{A}) \nonumber  \\
    & = \mathcal{A}^{y_{k}}(1-\mathcal{A})^{1-y_k}
\end{align}
If the total number of success (correct predictions) of a Bernoulli trial $y_1,y_2,\ldots,y_n$ is $c$, then it follows a Binomial distribution. 
\begin{align}
    p(c|\mathcal{A},n) &= B(c|\mathcal{A},n)\nonumber \\
    & = \binom{n}{c}\mathcal{A}^{n} (1-\mathcal{A})^{n-c}
\end{align}
This suggests choosing Beta density as the prior of $\mathcal{A}$ since it is the conjugate prior of the Binomial distribution. This implies 
\begin{align}
    p(\mathcal{A}) &= \text{Beta}(\mathcal{A}|a,b) \nonumber \\
    & = \text{Beta}(\mathcal{A}|1,1)
\end{align}
Now the posterior can be written using Bayes theorem as 
\begin{align}
    p(\mathcal{A}|c,n) & = \frac{p(c|\mathcal{A},n) p(\mathcal{A})}{p(c)} \nonumber \\
    & = \frac{B(c|\mathcal{A},n) \times \text{Beta}(\mathcal{A}|1,1)}{p(c)} \label{eqPosterior}
\end{align}
From equation \ref{eqPosterior}, we obtain the posterior $p(\mathcal{A}|c,n) = \text{Beta}(\mathcal{A}|c+1,n-c+1)$ and the posterior $(1-\alpha)100\%$ credible interval is \cite{carrillo2014probabilistic}
\begin{align}\label{credIntv}
    \Big[ F_{\text{Beta}(c+1,n-c+1)}^{-1} (\alpha/2); F_{\text{Beta}(c+1,n-c+1)}^{-1} (1-\alpha/2)\Big]
\end{align}
where $F_{\text{Beta}(\cdot)}^{-1} (\cdot)$ is the inverse density function of the Beta distribution and for 95\% credible interval, $\alpha=0.05$. In a multiclass scenario, each class has the distribution shown in equation (\ref{eqPosterior}). To find the posterior of the balanced accuracy $m-$fold convolution is used for $m$ classes. Numerical approximations are used to compute the posterior since analytical forms are not available for the $m-$fold convolution. In this work we have used a \textsc{Matlab} routine to compute the credible intervals of balanced accuracy provided in \cite{brodersen2010balanced}. 

\subsection{F1 measure}   
Another alternative performance evaluation metric is the F1-measure which has been used in some papers using the DEAP dataset \cite{koelstra2012deap,daimi2014classification,soleymani2017toolbox}. The F-measure was originally proposed by Van Rijsbergen \cite{van1979information} and is defined as \cite{chinchor1992muc}
\begin{equation}\label{fmeasure}
    F_{\beta} = \frac{(\beta^2+1)PR}{\beta^2P+R}
\end{equation}
where $P$ and $R$ denotes precision and recall and are defined as $P=tp/(tp+fp)$, $R=tp/(tp+fn)$ ($tp\rightarrow$ true positive, $fp\rightarrow$ false positive, $fn\rightarrow$ false negative). $\beta$ is a parameter to control balance between $P$ and $R$. When $\beta =1$, $F_1$ becomes the harmonic mean of precision and recall. Hence the $F_1$ measure is 
\begin{equation}\label{f1score}
     F_{1} = \frac{2PR}{P+R}
\end{equation}
Since $P$ and $R$ are calculated considering one class as a positive class, P and R have to be calculated per class and hence the F1 measure as well.  P and R per class can be calculated in two ways: microaveraging and macroaveraging. Microaveraging aggregates the individual true positives, false positives, and false negatives of each classes to calculate the $P$ and $R$.
\begin{align}
    &miP = \frac{\sum\limits_{k=1}^m C_{kk}}{\sum\limits_{k=1}^m C_{kk}+ \sum\limits_{k=1}^m \sum\limits_{\substack{j=1 \\ j\neq k}}^m C_{jk}}  \nonumber \\
    &miR = \frac{\sum\limits_{k=1}^m C_{kk}}{\sum\limits_{k=1}^m C_{kk}+ \sum\limits_{k=1}^m \sum\limits_{\substack{j=1 \\ j\neq k}}^m C_{kj}} \nonumber \\ 
    &miF_1 = \frac{2 \cdot miP \cdot miR}{miP +miR} \label{miF1}
\end{align}
An alternative technique is known as macroaveraging.  In macroaveraging, $P$ and $R$ are calculated for each classes and then $F_1$ for each class is computed using $P$ and $R$ of individual classes, and finally the macroaverage is the simple average of individual class $F_1$ scores.
\begin{align}
    &P_k = \frac{C_{kk} }{ C_{kk}+ \sum\limits_{\substack{j=0 \\ j\neq k}}^m C_{jk}} = \frac{C_{kk} }{\sum\limits_{j=1}^m C_{jk}}  \nonumber \\
    &R_k = \frac{ C_{kk}}{C_{kk}+ \sum\limits_{\substack{j=0 \\ j\neq k}}^m C_{kj}} = \frac{C_{kk} }{\sum\limits_{j=1}^m C_{kj}} \nonumber \\
    &maF_1 = \frac{1}{m} \sum_{k=1}^m \frac{2 \cdot P_k \cdot R_k}{P_k +R_k} \label{maF1}
\end{align}
The difference between $miF_1$ and $maF_1$ can be significant. Macro-averaging gives equal weight to each class, whereas micro-averaging gives equal weight to each per-class classification decision. Since $F_1$ measure ignores true negatives, the influence of large classes is higher than small classes in micro-averaging \cite{manning2010introduction}. However, the $F_1$ measure's harmonic means suggest that the averaging should be over the per-class classification decision of each instances. And in that case macro-averaging is not consistent with the original definition of the $F_1$ measure \cite{powers2015f}. Hence we yet do not have a convincing argument for choosing between $miF_1$ and $maF_1$  for multiclass classification. 

\section{Results}\label{res}

\begin{figure*}[ht!]
 \centering
     \subfigure[Accuracy and balanced accuracy]{
         \includegraphics[width=0.48\textwidth]{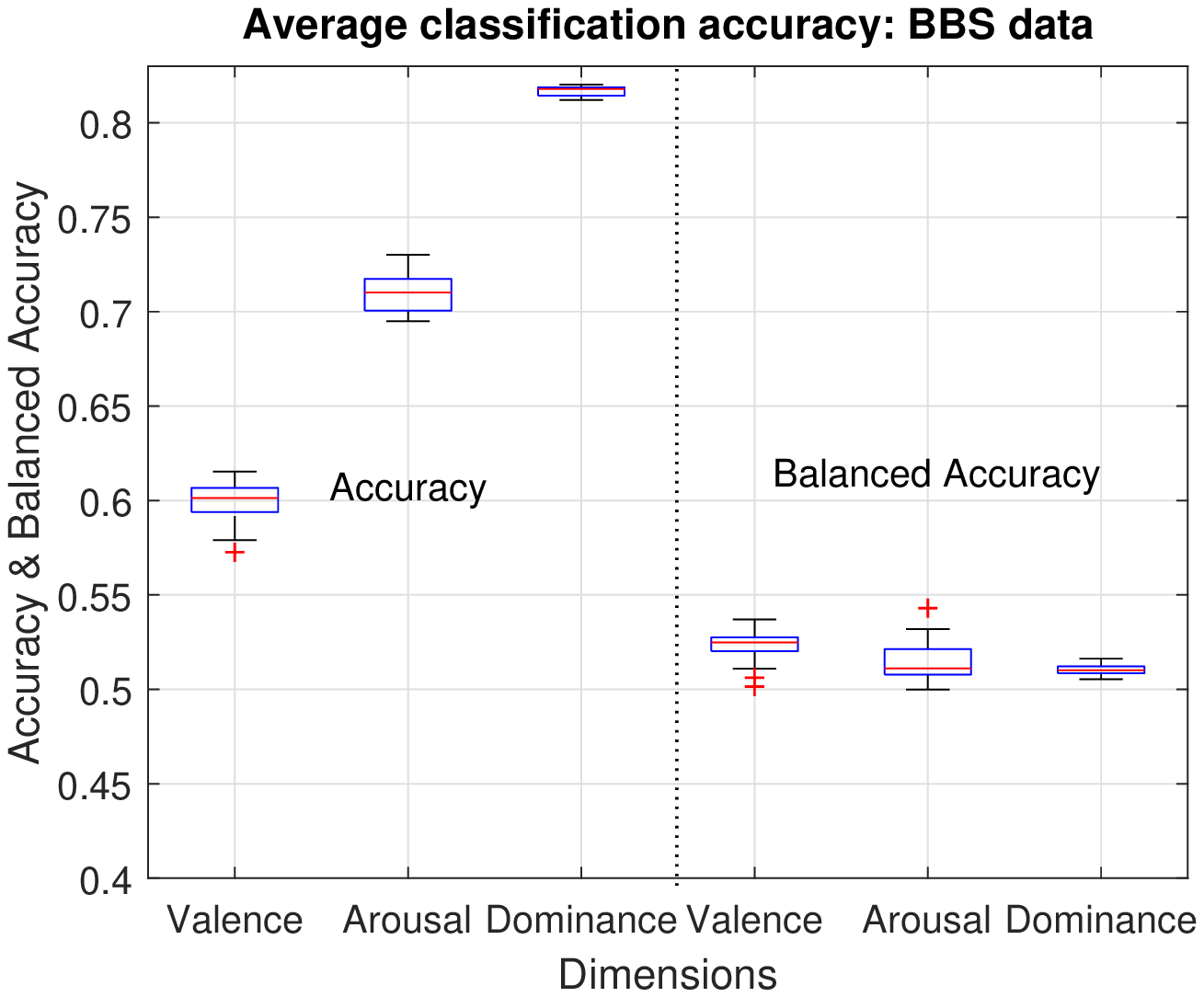} \label{baccbbs}}
    \subfigure[Macro and micro f1 measure]{
         \includegraphics[width=0.48\textwidth]{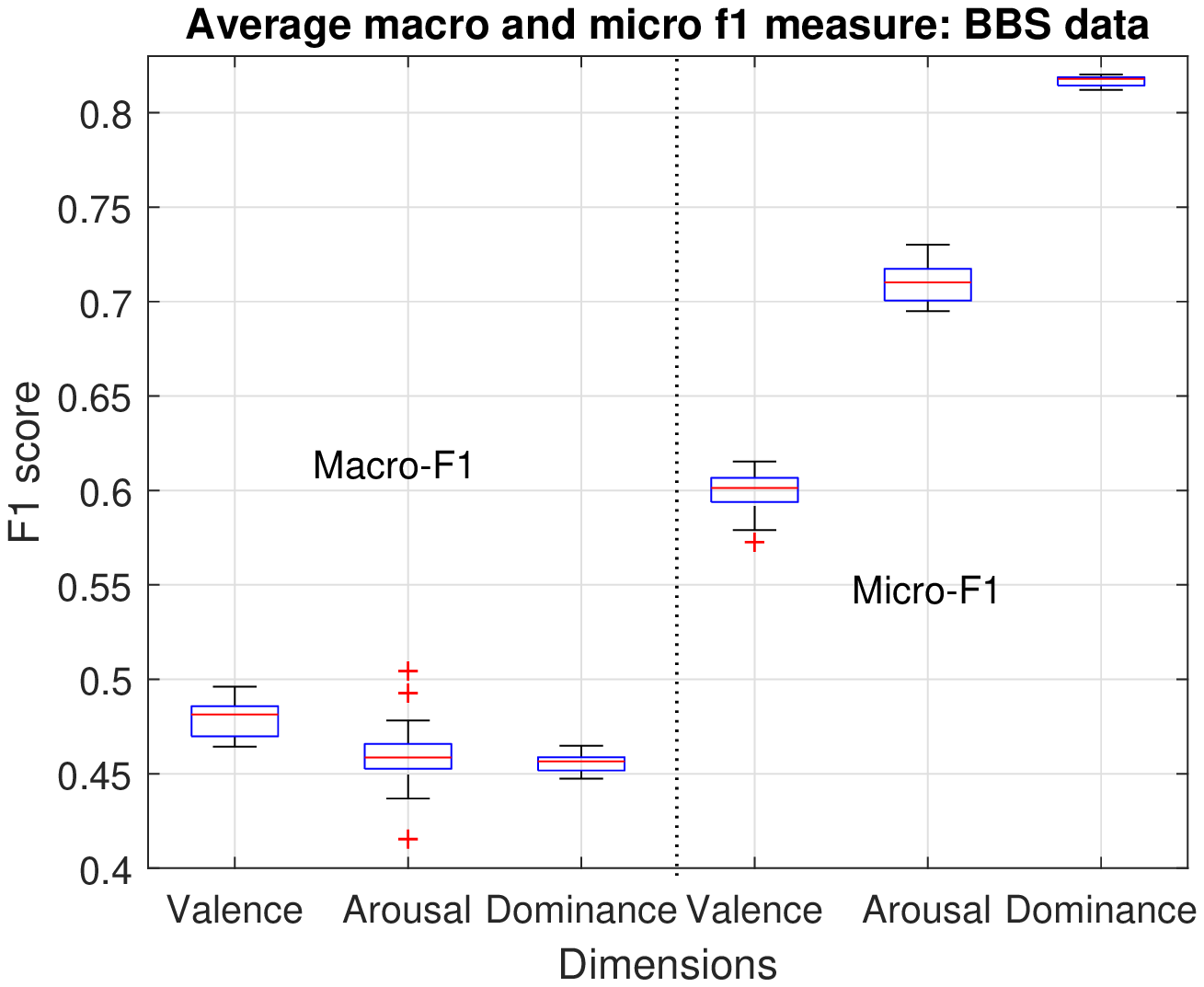}\label{f1bbs} }
     \caption{Average classification rate of all participants in valence, arousal and dominance recognition for different features using BBS data.}
     \label{fig:5}
\end{figure*}

 Since we have used seventeen different feature sets, it is not feasible to show all results here. To summarize the results, the classification results are averaged over all participants for each feature set. Those average classification accuracies, and other performance metrics for different feature sets, are presented in figure \ref{fig:4}, figure \ref{fig:5} and table \ref{table:2}. All the results presented here are for the SVM classifier since it performed better than the $k$NN approach. 

\subsection{DEAP Dataset}
\begin{figure*}[ht!]
 \centering
     \subfigure[Accuracy and balanced accuracy]{
         \includegraphics[width=0.48\textwidth]{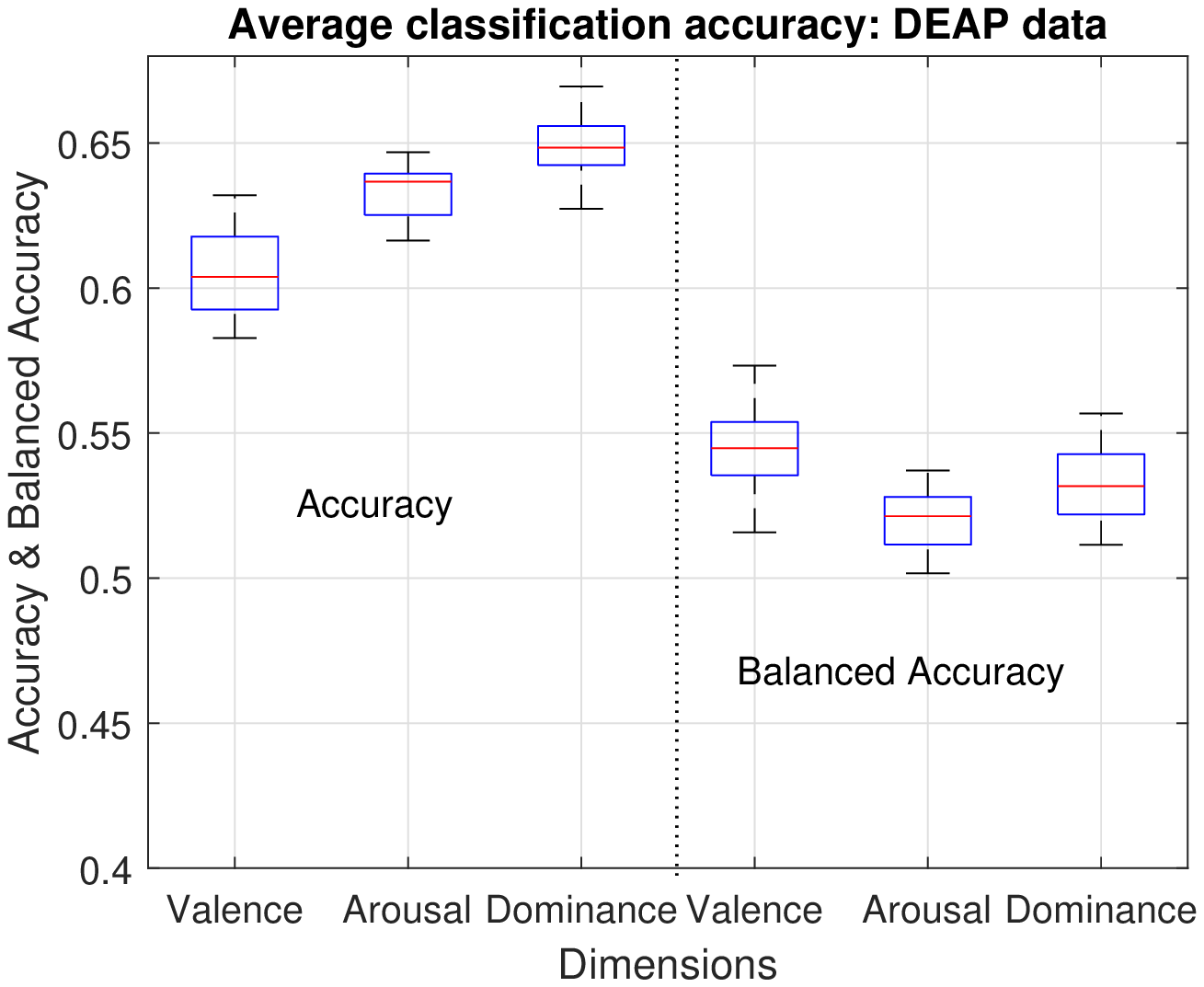} \label{baccdeap} }
    \subfigure[Macro and micro f1 measure]{
         \includegraphics[width=0.48\textwidth]{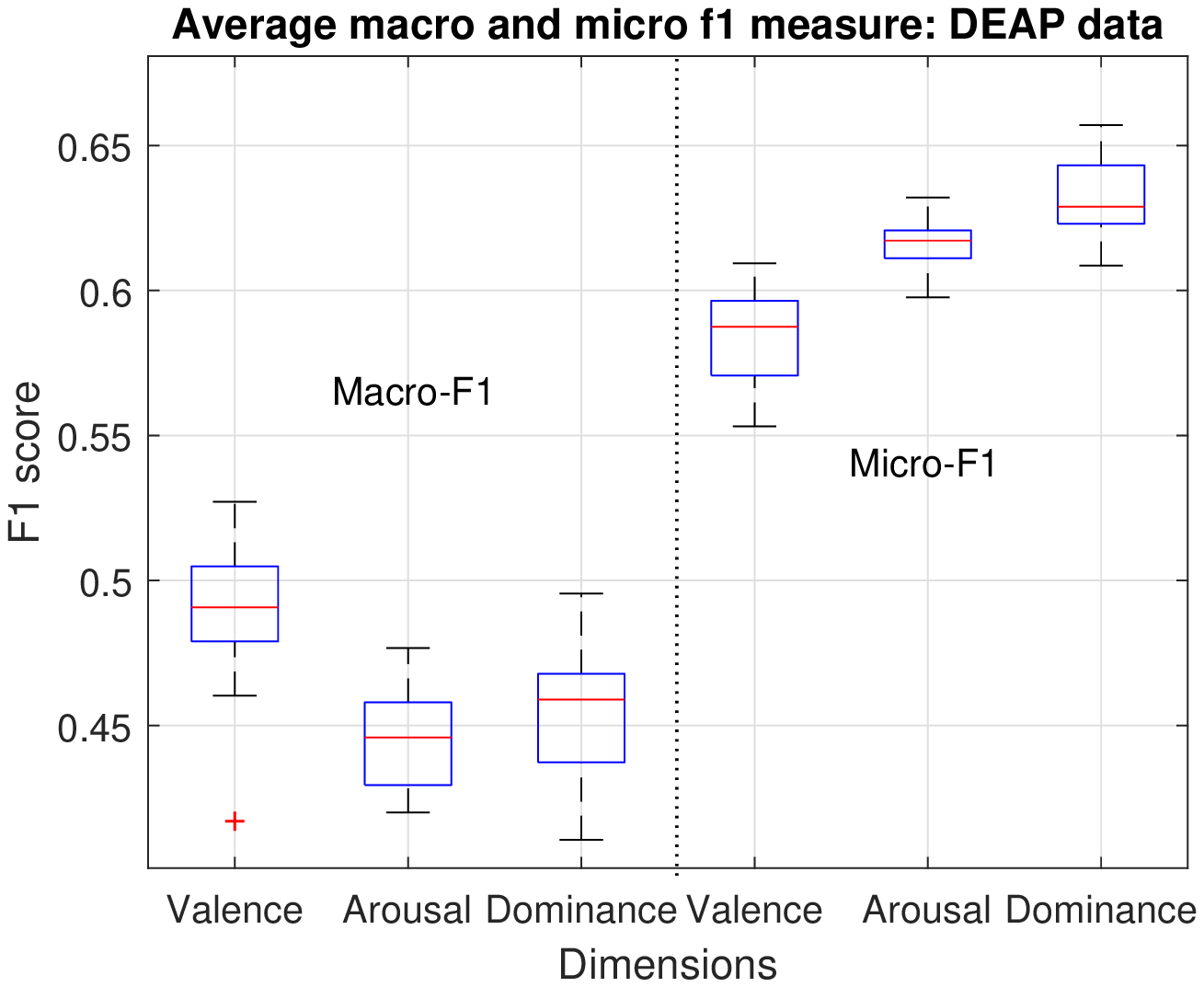}\label{f1deap} }
     \caption{Average classification rate of all participants in valence, arousal and dominance recognition for different features using DEAP dataset.}
     \label{fig:4}
\end{figure*}
Figure \ref{fig:4}(a) shows the average classification accuracies and balanced accuracies on DEAP for different feature sets using SVM. The mean classification rates for all features are 0.604, 0.637, and 0.648 for valence, arousal, and dominance, respectively. These results are very comparable with the results reported in DEAP original work \cite{koelstra2012deap} and other related studies \cite{soleymani2017toolbox,daimi2014classification}. But then if we check the balanced accuracies on the right side of the figure \ref{fig:4}(a), we will observe very different results. The mean classification rate in balanced accuracies for all feature are 0.544, 0.521, and 0.531 for valence, arousal and dominance respectively. These results are very different than the results with the simple accuracy metric except for valence recognition. The average class bias rate in these three dimensions are 0.59, 0.64 and 0.66 for valence, arousal, and dominance. 

Figure \ref{fig:4}(b) shows the average macro- and micro-averaged f1 measure for different feature sets using SVM. The mean macro-f1 for all feature are 0.49, 0.45 and 0.46 for valence, arousal, and dominance, respectively. On the contrary, the mean micro-f1 for all feature are 0.59, 0.62 and 0.63 for valence, arousal, and dominance, respectively. The best classification rate in the valence dimension is achieved using beta band power as a feature, as we found using balanced accuracy. For valence, the average across all participants macro-f1 for BetaP feature is 0.53 and the micro-f1 is 0.61. For arousal, the average across all participants macro-f1 for ThetaP feature is 0.48 and the micro-f1 is 0.63. For dominance, the average across all participants' macro-f1 for the TBR1 feature is $0.50$ and the micro-f1 is 0.65. 

\begin{table*}[ht!] 
\centering
\renewcommand{\arraystretch}{1.3}
\caption{The average for all participants classification rate in terms of balanced accuracy (BAcc) and the lower bound of the 95\% credible intervals of balanced accuracies for different feature sets.}  \label{table:2}
\setlength{\tabcolsep}{3pt}
\resizebox{0.8\textwidth}{!}{  
\begin{tabular}{l |cc|cc|cc}
\hline
\textbf{Features} & \multicolumn{2}{c|}{\textbf{Valence}} & \multicolumn{2}{c|}{\textbf{Arousal}} & \multicolumn{2}{c}{\textbf{Dominance}} \\
& Balanced & Lower bound & Balanced & Lower bound & Balanced & Lower bound \\
  & Accuracy (BAcc) & of BAcc & Accuracy (BAcc) & of BAcc & Accuracy (BAcc) & of BAcc \\
\hline
PASI & 0.5448 & 0.4297 & 0.5279 & 0.4227 & 0.5317 & 0.4262 \\
\hline                                                      
FAI & 0.5222 & 0.4089 & 0.5118 & 0.4130 & 0.5219 & 0.4178 \\ \hline                                                       TBR1 & 0.5479 & 0.4267 & 0.5247 & 0.4235 & \textbf{0.5568} & \textbf{0.4435} \\   
\hline                                                       TBR2 & 0.5381 & 0.4198 & 0.5109 & 0.4090 & 0.5220 & 0.4142 \\ \hline                                                       ThetaP & 0.5388 & 0.4211 & \textbf{0.5371} & \textbf{0.4336} & 0.5302 & 0.4206 \\ 
\hline                                                       AlphaP & 0.5432 & 0.4286 & 0.5238 & 0.4281 & 0.5492 & 0.4432 \\  \hline                                                 BetaP & \textbf{0.5732} & \textbf{0.4531} & 0.5303 & 0.4263 & 0.5370 & 0.4247 \\  
\hline                                          
GammaP & 0.5585 & 0.4381 & 0.5282 & 0.4265 & 0.5409 & 0.4323 \\  \hline                                                 TBR-C & 0.5663 & 0.4482 & 0.5318 & 0.4263 & 0.5550 & 0.4439 \\   \hline                                      
TABG & 0.5578 & 0.4401 & 0.5090 & 0.4122 & 0.5349 & 0.4301 \\ \hline                                                       
Hjorth & 0.5323 & 0.4159 & 0.5268 & 0.4268 & 0.5204 & 0.4104 \\ \hline                                                  
PASI+FASI & 0.5473 & 0.4355 & 0.5214 & 0.4207 & 0.5338 & 0.4307 \\
\hline                                                      
Avg-Entropy & 0.5158 & 0.4177 & 0.5200 & 0.4312 & 0.5176 & 0.4269 \\
\hline     
PSD & 0.5525 & 0.4451 & 0.5148 & 0.4259 & 0.5292 & 0.4361 \\ 
\hline
BARatio & 0.5230 & 0.4077 & 0.5016 & 0.4054 & 0.5115 & 0.4059 \\ \hline       
All & 0.5517 & 0.4447 & 0.5069 & 0.4178 & 0.5229 & 0.4290 \\
\hline                                     
All-PCA & 0.5365 & 0.4160 & 0.5166 & 0.4086 & 0.5484 & 0.4329 \\
\hline
\end{tabular}}
\end{table*}
Table \ref{table:2} shows the average balanced accuracies and lower bound of the 95\% credible intervals of balanced accuracies for different feature sets using equation (\ref{credIntv}). All results are for the SVM classifier. The highest obtained balanced accuracy across all dimensions is $0.5732$, achieved for valence recognition using beta band power. Unfortunately, the average lower limit of the credible intervals, in this case, is not above 0.5 (random chance). Though the average provides an overall recognition rate, it does not reflect the performance of individual participants. Explaining results for all features would be cumbersome; here we will explain classification results for each participant for only the best feature in each dimension. For valence, beta band power worked best. Using this feature, the balanced accuracy obtained for a participant (s10) with 0.75 and the lower bound of the credible interval is 0.622, which means that the valence classification rate is significantly above chance for this participant. Out of 32 participants, balanced accuracy is greater than 0.5 for 23 participants. For 8 of these participants, the lower bound of the credible interval is greater than 0.5. For arousal, theta band power worked best. Using the thetaP feature, the highest balanced accuracy obtained for a participant (s17) is 0.73 and the lower bound of the credible interval is 0.60, which means the arousal classification rate is significantly above chance for this participant. For 21 participants, observed balanced accuracy is greater than 0.5. However, only 4 participants were the lower bound of the credible interval greater than 0.5. For dominance, theta beta-1 ratio worked best. Using TBR1, the highest balanced accuracy obtained for a participant (s17) was 0.74 with a lower bound of 0.61, which means the dominance classification rate is significantly above chance for this participant. For 24 participants, balanced accuracy is greater than 0.5. Yet again, only for 4 participants was the lower bound of the credible interval greater than 0.5.

Table \ref{table:3} shows the affect recognition rate in terms of balanced accuracy, micro and macro averaged F1 score and also compared with the original work \cite{koelstra2012deap} and some other related studies. Rather than presenting the best results in each dimension, we chose to present results for one specific feature set for consistency. The results presented under the current study are for beta band power (BetaP) feature using an SVM classifier. Note that our comparison studies seem to have picked the best result in each dimension for their reported results (only Clerico et al. \cite{clerico2018eeg} unambiguously stated this).

\begin{table}[ht!] 
\renewcommand{\arraystretch}{1.3}
\caption{The classification rate in terms of balanced accuracy and micro F1 (miF1) and macro F1 (maF1) scores of affect recognition compared to the DEAP dataset original work and related studies. The results shown here are average of all participants for beta band power (BetaP) features.} 
\label{table:3}
\setlength{\tabcolsep}{3pt}
\resizebox{0.48\textwidth}{!}{  
\begin{tabular}{l ccc|ccc|ccc}
\hline
& \multicolumn{3}{c|}{\textbf{Valence}} & \multicolumn{3}{c|}{\textbf{Arousal}} 
& \multicolumn{3}{c}{\textbf{Dominance}} \\
&	 bAcc & miF1 & maF1 &  bAcc & miF1 & maF1 & bAcc & miF1 & maF1 \\   
\hline
\text{Koelstra et al. \cite{koelstra2012deap} }& --	& -- & 0.563 & -- & --& 0.583 & -- & -- & --	\\
\text{Daimi et al. \cite{daimi2014classification}}& --	& -- & 0.550&--	& --& 0.570 &--	&--& 0.552	\\
\text{Soleymani et al. \cite{soleymani2017toolbox} }& -- & -- & 0.645 & --	& -- & 0.570 & --	& -- & 0.533	\\
\text{Clerico et al. \cite{clerico2018eeg}}& 0.604	&--&--&	0.583	&--&--&		0.564& -- & --	\\
\text{Current study} &  0.573	& 0.610 &0.530 & 0.530	& 0.620& 0.460 &	0.537 & 0.630 & 0.460	\\
\hline
\end{tabular}}
\end{table}

\subsection{Data at BBS lab}
The data collected at the BBS lab using IAPS came from seven participants. For 2-class classification, the average class-bias were 0.60, 0.72, and 0.82 for valence, arousal, and dominance, respectively. For valence with SVM, the best 2-class classification results were obtained using gamma-band power considering the average of all participants. The obtained accuracy was 0.62 and the balanced accuracy was 0.54. The macro and micro averaged f1 scores were 0.49 and 0.60, respectively.

For arousal with SVM, the best 2-class classification results were obtained using the power asymmetry index (PASI) considering the average of all participants. The obtained accuracy was 0.73 and the balanced accuracy was 0.54. The macro and micro averaged f1 scores were 0.50 and 0.71, respectively. 

For dominance with SVM, the best 2-class classification results were obtained using beta band power considering the average of all participants. The obtained accuracy was 0.82 and the balanced accuracy was 0.52. The macro and micro averaged f1 scores were 0.46 and 0.82, respectively.

\section{Discussion}\label{diss}
For the DEAP, the average class bias or majority class percentage in a 2-class classification scenario for valence, arousal and dominance are 0.59, 0.64 and 0.66 respectively. We have argued that class imbalance is important to understand the results of the classifier and should be reported. Performance metrics that include or account the class-biases are thus preferred to use. Any metric that ignores class imbalance will mislead readers. To illustrate this, consider the results from table \ref{table:2} where balanced accuracies and its lower bound of the 95\% credible interval were presented for different feature sets for DEAP data using SVM. The best average classification accuracy for all participants in the valence dimension was 0.602 using beta band power as a feature, whereas the balanced accuracy, in this case, was 0.573. Without knowing the class bias and considering the accuracy metric, one might think the result is promising. But the lower bound of the 95\% credible interval of balanced accuracy shows that the classification rate can not be claimed as statistically significant.

However, class imbalance for each participant for all three-dimension (valence, arousal, dominance) would be cumbersome and impractical to report. The biases mentioned earlier were averaged across all participants. Since affective state estimation is a participant-specific task, averaged results do not reflect individual performances. So comparisons using average results are not meaningful. Hence, we need something else which can address both the class imbalance problem and make the average performance meaningful. Considering those above-mentioned problems, balanced accuracy is a promising candidate since the baseline performance for balance accuracy is the same (50\%) across all dimensions(valence, arousal, dominance) for all participants. Thus, balanced accuracy will make results easier to understand and compare. For example, just looking at the results in table \ref{table:2}, we can easily conclude that the valence recognition rate is better than arousal and dominance recognition. Statistical comparison between the balanced accuracies for valence, arousal and dominance presented in table \ref{table:2} is done by using \textsc{MATLAB} inbuilt function \texttt{ttest2}. Two-sample t-test resulted in the rejection of the null hypothesis (two groups are equal) when comparing valence and arousal. The valence recognition rate is significantly better than the arousal and dominance recognition rate with $p$-values $0.035$ and $7.44e^{-06}$. The dominance recognition rate is also significantly better than arousal with $p$-value of $0.031$. These three two-sample t-tests suggested that valence has the highest recognition rate and arousal has the lowest for the DEAP dataset.

Averages for all participants of the balanced accuracies, macro, and micro f1 measure were compared with other related studies in Table \ref{table:3}. Since they have not discussed the methods of statistical analysis, here we will use our obtained results shown in table \ref{table:2} for discussion. Our average balanced accuracies are very similar to the highest balanced accuracy reported in \cite{clerico2018eeg}. In \cite{clerico2018eeg}, it has been claimed that all the reported balanced accuracies were better than random voting classifiers with $p<0.05$. This statement is true if we perform statistical analysis considering results from all participants as a group rather than individual participants. The number of participants with balanced accuracy above $0.5$ is $25$ for valence using all frequency band powers, $21$ for arousal and 20 for dominance. In this case the probability that overall balanced accuracy is above chance are $0.66$, $0.66$ and $0.63$ with intervals $(0.47-0.82)$,  $(0.47-0.82)$, and $(0.44-0.79)$ for valence, arousal and dominance, respectively. But the significance of the experiment as a whole does not capture the significance of each participant's performance. Hence, just based on these statistics we are not comfortable to claim the accuracies are above chance. Rather we suggest using the probability of individual participants' performances being above chance to claim the results are significant. Using the number of participants that are significantly above chance, we have 6 for valence, 3 for arousal and 4 for dominance out of 32 participants. That tells us that the probabilities of a participant's  classification accuracy being significantly above chance for valence, arousal and dominance are $0.19$, $0.09$ and $0.13$ bounded by $(0.07-0.36)$, $(0.02-0.25)$ and $(0.04-.29)$, respectively. These are not very encouraging, as valence is only above the typical 0.05 threshold. This low rate of significant performance may be of concern for the EEG based affective computing community, and as a community, we need to be more careful while reporting results.

\section{Conclusion}\label{cons}
In this work,  we presented the experimental results for affective state estimation using the publicly available DEAP database and our lab data. We compared our results for DEAP data with the results reported in other few related studies. We used various features mentioned in the literature and also investigated theta-beta1 ratio as a novel features for affect classification. Our findings showed that the beta band power is most suitable for valence classification, theta band power for arousal classification, and theta beta-1 ratio for dominance classification. 

In conclusion, we suggest using balanced accuracy and its posterior distribution as the performance evaluation metric for emotion estimation. Though F1 measure is a popular choice, it is not yet well established which F1 measure (macro/micro) we should use for multiclass classification.  As our results demonstrate, that choice is important. Further, if macro-averaging is chosen, the statistical significance of the metric is not well understood.

In contrast to the F1 measure, balanced accuracy has several advantages.  First, balanced accuracy does not have a "preferred class" and is thus comparable between groups.  Second, the credible bounds can be calculated using known formulas. Third, the extension to large numbers of classes is straightforward. Fourth and finally, balanced accuracy is insensitive to class bias and always has the intuitive 1/k chance performance for unskilled classifiers. 

We note that traditional accuracy metrics would have classified the performance of many more of our participants as statistically significant, relative to the number classified this way by balanced accuracy.  Nevertheless, we maintain that balanced accuracy is far less misleading, and that the traditional accuracy metric substantially over-estimates performance is these unbalanced datasets. \\
\\

\textsc{Acknowledgment}
\par The authors would like to thank our participants for enduring long EEG sessions. Opinions, findings, conclusions, or recommendations expressed in this material are those of the authors and do not necessarily reflect the views of the funding agencies. The involvement of human participants with this research was approved by the Kansas State University Institutional Review Board under protocol No. 8328. \\

\bibliographystyle{IEEEtran}
\bibliography{references}

\begin{thebibliography}{10}
\providecommand{\url}[1]{#1}
\csname url@samestyle\endcsname
\providecommand{\newblock}{\relax}
\providecommand{\bibinfo}[2]{#2}
\providecommand{\BIBentrySTDinterwordspacing}{\spaceskip=0pt\relax}
\providecommand{\BIBentryALTinterwordstretchfactor}{4}
\providecommand{\BIBentryALTinterwordspacing}{\spaceskip=\fontdimen2\font plus
\BIBentryALTinterwordstretchfactor\fontdimen3\font minus
  \fontdimen4\font\relax}
\providecommand{\BIBforeignlanguage}[2]{{%
\expandafter\ifx\csname l@#1\endcsname\relax
\typeout{** WARNING: IEEEtran.bst: No hyphenation pattern has been}%
\typeout{** loaded for the language `#1'. Using the pattern for}%
\typeout{** the default language instead.}%
\else
\language=\csname l@#1\endcsname
\fi
#2}}
\providecommand{\BIBdecl}{\relax}
\BIBdecl

\bibitem{wolpaw2000brain}
J.~R. Wolpaw, N.~Birbaumer, W.~J. Heetderks, D.~J. McFarland, P.~H. Peckham,
  G.~Schalk, E.~Donchin, L.~A. Quatrano, C.~J. Robinson, T.~M. Vaughan
  \emph{et~al.}, ``Brain-computer interface technology: a review of the first
  international meeting,'' \emph{IEEE transactions on rehabilitation
  engineering}, vol.~8, no.~2, pp. 164--173, 2000.

\bibitem{forgas1995mood}
J.~P. Forgas, ``Mood and judgment: the affect infusion model (aim).''
  \emph{Psychological bulletin}, vol. 117, no.~1, p.~39, 1995.

\bibitem{salovey1990emotional}
P.~Salovey and J.~D. Mayer, ``Emotional intelligence,'' \emph{Imagination,
  cognition and personality}, vol.~9, no.~3, pp. 185--211, 1990.

\bibitem{goleman1996emotional}
D.~Goleman, ``Emotional intelligence. {W}hy it can matter more than {IQ}.''
  \emph{Learning}, vol.~24, no.~6, pp. 49--50, 1996.

\bibitem{strack1989salience}
F.~Strack, N.~Schwarz, B.~Chassein, D.~Kern, and D.~Wagner, ``Salience of
  comparison standards and the activation of social norms: Consequences for
  judgements of happiness and their communication,'' \emph{British Journal of
  Social Psychology}, vol.~29, no.~4, pp. 303--314, 1990.

\bibitem{picard2001toward}
R.~W. Picard, E.~Vyzas, and J.~Healey, ``Toward machine emotional intelligence:
  Analysis of affective physiological state,'' \emph{IEEE transactions on
  pattern analysis and machine intelligence}, vol.~23, no.~10, pp. 1175--1191,
  2001.

\bibitem{pantic2000automatic}
M.~Pantic and L.~J.~M. Rothkrantz, ``Automatic analysis of facial expressions:
  The state of the art,'' \emph{IEEE Transactions on pattern analysis and
  machine intelligence}, vol.~22, no.~12, pp. 1424--1445, 2000.

\bibitem{niemic2002studies}
C.~P. Niemic and K.~Warren, ``Studies of emotion: A theoretical and empirical
  review of psychophysiological studies of emotion,'' \emph{Journal of
  Undergraduate Research Rochester}, vol.~1, no.~1, pp. 15--19, 2002.

\bibitem{russell1980circumplex}
J.~A. Russell, ``A circumplex model of affect.'' \emph{Journal of personality
  and social psychology}, vol.~39, no.~6, p. 1161, 1980.

\bibitem{cacioppo1990inferring}
J.~T. Cacioppo and L.~G. Tassinary, ``Inferring psychological significance from
  physiological signals.'' \emph{American psychologist}, vol.~45, no.~1, p.~16,
  1990.

\bibitem{koelstra2012deap}
S.~Koelstra, C.~Muhl, M.~Soleymani, J.-S. Lee, A.~Yazdani, T.~Ebrahimi, T.~Pun,
  A.~Nijholt, and I.~Patras, ``Deap: A database for emotion analysis; using
  physiological signals,'' \emph{IEEE Transactions on Affective Computing},
  vol.~3, no.~1, pp. 18--31, 2012.

\bibitem{lang2008iaps}
P.~J. Lang, M.~M. Bradley, and B.~N. Cuthbert, ``International affective
  picture system ({IAPS}): Affective ratings of pictures and instruction
  manual,'' The Center for Research in Psychophysiology, University of Florida,
  FL, USA, Technical Report A-8, 2008.

\bibitem{lang1999iads}
P.~J. Lang and M.~M. Bradley, ``International affective digitized sounds
  ({IADS}): Stimuli, instruction manual and affective ratings,'' The Center for
  Research in Psychophysiology, University of Florida, FL, USA, Technical
  Report B-2, 1999.

\bibitem{baveye2015liris}
Y.~Baveye, E.~Dellandrea, C.~Chamaret, and L.~Chen, ``Liris-accede: A video
  database for affective content analysis,'' \emph{IEEE Transactions on
  Affective Computing}, vol.~6, no.~1, pp. 43--55, 2015.

\bibitem{guntekin2014review}
B.~G{\"u}ntekin and E.~Ba{\c{s}}ar, ``A review of brain oscillations in
  perception of faces and emotional pictures,'' \emph{Neuropsychologia},
  vol.~58, pp. 33--51, 2014.

\bibitem{wagh2019electroencephalograph}
K.~P. Wagh and K.~Vasanth, ``Electroencephalograph (eeg) based emotion
  recognition system: A review,'' in \emph{Innovations in Electronics and
  Communication Engineering}.\hskip 1em plus 0.5em minus 0.4em\relax Springer,
  2019, pp. 37--59.

\bibitem{garcia2019review}
B.~Garc{\'\i}a-Mart{\'\i}nez, A.~Martinez-Rodrigo, R.~Alcaraz, and
  A.~Fern{\'a}ndez-Caballero, ``A review on nonlinear methods using
  electroencephalographic recordings for emotion recognition,'' \emph{IEEE
  Transactions on Affective Computing}, 2019.

\bibitem{wang2014hybrid}
S.~Wang, Y.~Zhu, G.~Wu, and Q.~Ji, ``Hybrid video emotional tagging using
  users’ eeg and video content,'' \emph{Multimedia tools and applications},
  vol.~72, no.~2, pp. 1257--1283, 2014.

\bibitem{soleymani2011multimodal}
M.~Soleymani, J.~Lichtenauer, T.~Pun, and M.~Pantic, ``A multimodal database
  for affect recognition and implicit tagging,'' \emph{IEEE Transactions on
  Affective Computing}, vol.~3, no.~1, pp. 42--55, 2011.

\bibitem{piho2018mutual}
L.~Piho and T.~Tjahjadi, ``A mutual information based adaptive windowing of
  informative eeg for emotion recognition,'' \emph{IEEE Transactions on
  Affective Computing}, 2018.

\bibitem{li2018exploring}
X.~Li, D.~Song, P.~Zhang, Y.~Zhang, Y.~Hou, and B.~Hu, ``Exploring eeg features
  in cross-subject emotion recognition,'' \emph{Frontiers in neuroscience},
  vol.~12, p. 162, 2018.

\bibitem{soleymani2017toolbox}
M.~Soleymani, F.~Villaro-Dixon, T.~Pun, and G.~Chanel, ``Toolbox for emotional
  feature extraction from physiological signals (teap),'' \emph{Frontiers in
  ICT}, vol.~4, p.~1, 2017.

\bibitem{zheng2017identifying}
W.-L. Zheng, J.-Y. Zhu, and B.-L. Lu, ``Identifying stable patterns over time
  for emotion recognition from eeg,'' \emph{IEEE Transactions on Affective
  Computing}, 2017.

\bibitem{wang2017content}
S.~Wang, S.~Chen, and Q.~Ji, ``Content-based video emotion tagging augmented by
  users’ multiple physiological responses,'' \emph{IEEE Transactions on
  Affective Computing}, vol.~10, no.~2, pp. 155--166, 2017.

\bibitem{ozerdem2017emotion}
M.~S. {\"O}zerdem and H.~Polat, ``Emotion recognition based on eeg features in
  movie clips with channel selection,'' \emph{Brain informatics}, vol.~4,
  no.~4, p. 241, 2017.

\bibitem{verma2017affect}
G.~K. Verma and U.~S. Tiwary, ``Affect representation and recognition in 3d
  continuous valence--arousal--dominance space,'' \emph{Multimedia Tools and
  Applications}, vol.~76, no.~2, pp. 2159--2183, 2017.

\bibitem{li2017human}
Y.~Li, J.~Huang, H.~Zhou, and N.~Zhong, ``Human emotion recognition with
  electroencephalographic multidimensional features by hybrid deep neural
  networks,'' \emph{Applied Sciences}, vol.~7, no.~10, p. 1060, 2017.

\bibitem{garcia2016application}
B.~Garc{\'\i}a-Mart{\'\i}nez, A.~Mart{\'\i}nez-Rodrigo,
  R.~Zangr{\'o}niz~Cantabrana, J.~Pastor~Garc{\'\i}a, and R.~Alcaraz,
  ``Application of entropy-based metrics to identify emotional distress from
  electroencephalographic recordings,'' \emph{Entropy}, vol.~18, no.~6, p. 221,
  2016.

\bibitem{al2018anytime}
O.~Al~Zoubi, M.~Awad, and N.~K. Kasabov, ``Anytime multipurpose emotion
  recognition from eeg data using a liquid state machine based framework,''
  \emph{Artificial intelligence in medicine}, vol.~86, pp. 1--8, 2018.

\bibitem{purnamasari2017development}
P.~Purnamasari, A.~Ratna, and B.~Kusumoputro, ``Development of filtered
  bispectrum for eeg signal feature extraction in automatic emotion recognition
  using artificial neural networks,'' \emph{Algorithms}, vol.~10, no.~2, p.~63,
  2017.

\bibitem{liu2018emotion}
J.~Liu, H.~Meng, M.~Li, F.~Zhang, R.~Qin, and A.~K. Nandi, ``Emotion detection
  from eeg recordings based on supervised and unsupervised dimension
  reduction,'' \emph{Concurrency and Computation: Practice and Experience},
  vol.~30, no.~23, p. e4446, 2018.

\bibitem{torres2017svm}
C.~Torres-Valencia, M.~{\'A}lvarez-L{\'o}pez, and {\'A}.~Orozco-Guti{\'e}rrez,
  ``Svm-based feature selection methods for emotion recognition from multimodal
  data,'' \emph{Journal on Multimodal User Interfaces}, vol.~11, no.~1, pp.
  9--23, 2017.

\bibitem{menezes2017towards}
M.~L.~R. Menezes, A.~Samara, L.~Galway, A.~Sant’Anna, A.~Verikas,
  F.~Alonso-Fernandez, H.~Wang, and R.~Bond, ``Towards emotion recognition for
  virtual environments: an evaluation of eeg features on benchmark dataset,''
  \emph{Personal and Ubiquitous Computing}, vol.~21, no.~6, pp. 1003--1013,
  2017.

\bibitem{mert2018emotion}
A.~Mert and A.~Akan, ``Emotion recognition from eeg signals by using
  multivariate empirical mode decomposition,'' \emph{Pattern Analysis and
  Applications}, vol.~21, no.~1, pp. 81--89, 2018.

\bibitem{nakisa2018evolutionary}
B.~Nakisa, M.~N. Rastgoo, D.~Tjondronegoro, and V.~Chandran, ``Evolutionary
  computation algorithms for feature selection of eeg-based emotion recognition
  using mobile sensors,'' \emph{Expert Systems with Applications}, vol.~93, pp.
  143--155, 2018.

\bibitem{hemanth2018brain}
D.~J. Hemanth, J.~Anitha \emph{et~al.}, ``Brain signal based human emotion
  analysis by circular back propagation and deep kohonen neural networks,''
  \emph{Computers \& Electrical Engineering}, vol.~68, pp. 170--180, 2018.

\bibitem{yin2017cross}
Z.~Yin, Y.~Wang, L.~Liu, W.~Zhang, and J.~Zhang, ``Cross-subject eeg feature
  selection for emotion recognition using transfer recursive feature
  elimination,'' \emph{Frontiers in neurorobotics}, vol.~11, p.~19, 2017.

\bibitem{yoon2013eeg}
H.~J. Yoon and S.~Y. Chung, ``Eeg-based emotion estimation using bayesian
  weighted-log-posterior function and perceptron convergence algorithm,''
  \emph{Computers in biology and medicine}, vol.~43, no.~12, pp. 2230--2237,
  2013.

\bibitem{wang2015emotion}
S.~Wang, Y.~Zhu, L.~Yue, and Q.~Ji, ``Emotion recognition with the help of
  privileged information,'' \emph{IEEE Transactions on Autonomous Mental
  Development}, vol.~7, no.~3, pp. 189--200, 2015.

\bibitem{jirayucharoensak2014eeg}
S.~Jirayucharoensak, S.~Pan-Ngum, and P.~Israsena, ``Eeg-based emotion
  recognition using deep learning network with principal component based
  covariate shift adaptation,'' \emph{The Scientific World Journal}, vol. 2014,
  2014.

\bibitem{daimi2014classification}
S.~N. Daimi and G.~Saha, ``Classification of emotions induced by music videos
  and correlation with participants’ rating,'' \emph{Expert Systems with
  Applications}, vol.~41, no.~13, pp. 6057--6065, 2014.

\bibitem{padilla2016emotion}
J.~I. Padilla-Buritica, J.~D. Martinez-Vargas, and G.~Castellanos-Dominguez,
  ``Emotion discrimination using spatially compact regions of interest
  extracted from imaging eeg activity,'' \emph{Frontiers in computational
  neuroscience}, vol.~10, p.~55, 2016.

\bibitem{gupta2016relevance}
R.~Gupta, T.~H. Falk \emph{et~al.}, ``Relevance vector classifier decision
  fusion and eeg graph-theoretic features for automatic affective state
  characterization,'' \emph{Neurocomputing}, vol. 174, pp. 875--884, 2016.

\bibitem{yin2017recognition}
Z.~Yin, M.~Zhao, Y.~Wang, J.~Yang, and J.~Zhang, ``Recognition of emotions
  using multimodal physiological signals and an ensemble deep learning model,''
  \emph{Computer methods and programs in biomedicine}, vol. 140, pp. 93--110,
  2017.

\bibitem{clerico2018eeg}
A.~Clerico, A.~Tiwari, R.~Gupta, S.~Jayaraman, and T.~H. Falk,
  ``Electroencephalography amplitude modulation analysis for automated
  affective tagging of music video clips,'' \emph{Frontiers in computational
  neuroscience}, vol.~11, p. 115, 2018.

\bibitem{lobo2008auc}
J.~M. Lobo, A.~Jim{\'e}nez-Valverde, and R.~Real, ``Auc: a misleading measure
  of the performance of predictive distribution models,'' \emph{Global ecology
  and Biogeography}, vol.~17, no.~2, pp. 145--151, 2008.

\bibitem{van2013macro}
V.~Van~Asch, ``Macro-and micro-averaged evaluation measures [[basic draft]],''
  \emph{Belgium: CLiPS}, pp. 1--27, 2013.

\bibitem{bradley1994measuring}
M.~M. Bradley and P.~J. Lang, ``Measuring emotion: the self-assessment manikin
  and the semantic differential,'' \emph{Journal of behavior therapy and
  experimental psychiatry}, vol.~25, no.~1, pp. 49--59, 1994.

\bibitem{schalk2004bci2000}
G.~Schalk, D.~J. McFarland, T.~Hinterberger, N.~Birbaumer, and J.~R. Wolpaw,
  ``{BCI}2000: a general-purpose brain-computer interface ({BCI}) system,''
  \emph{IEEE Transactions on biomedical engineering}, vol.~51, no.~6, pp.
  1034--1043, 2004.

\bibitem{velo2012should}
J.~R. Velo, J.~L. Stewart, B.~P. Hasler, D.~N. Towers, and J.~J. Allen,
  ``Should it matter when we record? time of year and time of day as factors
  influencing frontal eeg asymmetry,'' \emph{Biological psychology}, vol.~91,
  no.~2, pp. 283--291, 2012.

\bibitem{allen2015frontal}
J.~J. Allen and S.~J. Reznik, ``Frontal eeg asymmetry as a promising marker of
  depression vulnerability: Summary and methodological considerations,''
  \emph{Current opinion in psychology}, vol.~4, pp. 93--97, 2015.

\bibitem{kayser2006principal}
J.~Kayser and C.~E. Tenke, ``Principal components analysis of laplacian
  waveforms as a generic method for identifying erp generator patterns: I.
  evaluation with auditory oddball tasks,'' \emph{Clinical neurophysiology},
  vol. 117, no.~2, pp. 348--368, 2006.

\bibitem{perrin1989spherical}
F.~Perrin, J.~Pernier, O.~Bertrand, and J.~Echallier, ``Spherical splines for
  scalp potential and current density mapping,'' \emph{Electroencephalography
  and clinical neurophysiology}, vol.~72, no.~2, pp. 184--187, 1989.

\bibitem{lin2010eeg}
Y.-P. Lin, C.-H. Wang, T.-P. Jung, T.-L. Wu, S.-K. Jeng, J.-R. Duann, and J.-H.
  Chen, ``Eeg-based emotion recognition in music listening,'' \emph{IEEE
  Transactions on Biomedical Engineering}, vol.~57, no.~7, pp. 1798--1806,
  2010.

\bibitem{jenke2014feature}
R.~Jenke, A.~Peer, and M.~Buss, ``Feature extraction and selection for emotion
  recognition from eeg,'' \emph{IEEE Transactions on Affective Computing},
  vol.~5, no.~3, pp. 327--339, 2014.

\bibitem{welch1967use}
P.~Welch, ``The use of fast fourier transform for the estimation of power
  spectra: a method based on time averaging over short, modified
  periodograms,'' \emph{IEEE Transactions on audio and electroacoustics},
  vol.~15, no.~2, pp. 70--73, 1967.

\bibitem{coan2004frontal}
J.~A. Coan and J.~J. Allen, ``Frontal eeg asymmetry as a moderator and mediator
  of emotion,'' \emph{Biological psychology}, vol.~67, no. 1-2, pp. 7--50,
  2004.

\bibitem{allen2004issues}
J.~J. Allen, J.~A. Coan, and M.~Nazarian, ``Issues and assumptions on the road
  from raw signals to metrics of frontal eeg asymmetry in emotion,''
  \emph{Biological psychology}, vol.~67, no. 1-2, pp. 183--218, 2004.

\bibitem{van2017frontal}
N.~van~der Vinne, M.~A. Vollebregt, M.~J. van Putten, and M.~Arns, ``Frontal
  alpha asymmetry as a diagnostic marker in depression: Fact or fiction? a
  meta-analysis,'' \emph{Neuroimage: clinical}, vol.~16, pp. 79--87, 2017.

\bibitem{putman2010eeg}
P.~Putman, J.~van Peer, I.~Maimari, and S.~van~der Werff, ``Eeg theta/beta
  ratio in relation to fear-modulated response-inhibition, attentional control,
  and affective traits,'' \emph{Biological psychology}, vol.~83, no.~2, pp.
  73--78, 2010.

\bibitem{hjorth1970eeg}
B.~Hjorth, ``Eeg analysis based on time domain properties,''
  \emph{Electroencephalography and clinical neurophysiology}, vol.~29, no.~3,
  pp. 306--310, 1970.

\bibitem{vakkuri2004time}
A.~Vakkuri, A.~Yli-Hankala, P.~Talja, S.~Mustola, H.~Tolvanen-Laakso,
  T.~Sampson, and H.~Vierti{\"o}-Oja, ``Time-frequency balanced spectral
  entropy as a measure of anesthetic drug effect in central nervous system
  during sevoflurane, propofol, and thiopental anesthesia,'' \emph{Acta
  Anaesthesiologica Scandinavica}, vol.~48, no.~2, pp. 145--153, 2004.

\bibitem{mohammadi2017wavelet}
Z.~Mohammadi, J.~Frounchi, and M.~Amiri, ``Wavelet-based emotion recognition
  system using eeg signal,'' \emph{Neural Computing and Applications}, vol.~28,
  no.~8, pp. 1985--1990, 2017.

\bibitem{murphy2012machine}
K.~P. Murphy, \emph{Machine Learning: A Probabilistic Perspective}, ser.
  Adaptive Computation and Machine Learning.\hskip 1em plus 0.5em minus
  0.4em\relax MIT Press, 2012.

\bibitem{lachiche2003improving}
N.~Lachiche and P.~A. Flach, ``Improving accuracy and cost of two-class and
  multi-class probabilistic classifiers using roc curves,'' in
  \emph{Proceedings of the 20th International Conference on Machine Learning
  (ICML-03)}, 2003, pp. 416--423.

\bibitem{sokolova2009systematic}
M.~Sokolova and G.~Lapalme, ``A systematic analysis of performance measures for
  classification tasks,'' \emph{Information Processing \& Management}, vol.~45,
  no.~4, pp. 427--437, 2009.

\bibitem{velez2007balanced}
D.~R. Velez, B.~C. White, A.~A. Motsinger, W.~S. Bush, M.~D. Ritchie, S.~M.
  Williams, and J.~H. Moore, ``A balanced accuracy function for epistasis
  modeling in imbalanced datasets using multifactor dimensionality reduction,''
  \emph{Genetic epidemiology}, vol.~31, no.~4, pp. 306--315, 2007.

\bibitem{brodersen2010balanced}
K.~H. Brodersen, C.~S. Ong, K.~E. Stephan, and J.~M. Buhmann, ``The balanced
  accuracy and its posterior distribution,'' in \emph{Pattern recognition
  (ICPR), 2010 20th international conference on}.\hskip 1em plus 0.5em minus
  0.4em\relax IEEE, 2010, pp. 3121--3124.

\bibitem{carrillo2014probabilistic}
H.~Carrillo, K.~H. Brodersen, and J.~A. Castellanos, ``Probabilistic
  performance evaluation for multiclass classification using the posterior
  balanced accuracy,'' in \emph{ROBOT2013: First Iberian Robotics
  Conference}.\hskip 1em plus 0.5em minus 0.4em\relax Springer, 2014, pp.
  347--361.

\bibitem{van1979information}
C.~J. Van~Rijsbergen, ``Information retrieval. 2nd. newton, ma,'' 1979.

\bibitem{chinchor1992muc}
N.~Chinchor, ``Muc-4 evaluation metrics,'' in \emph{Proceedings of the 4th
  conference on Message understanding}.\hskip 1em plus 0.5em minus 0.4em\relax
  Association for Computational Linguistics, 1992, pp. 22--29.

\bibitem{manning2010introduction}
C.~Manning, P.~Raghavan, and H.~Sch{\"u}tze, ``Introduction to information
  retrieval,'' \emph{Natural Language Engineering}, vol.~16, no.~1, pp.
  100--103, 2010.

\bibitem{powers2015f}
D.~M. Powers, ``What the f-measure doesn't measure: Features, flaws, fallacies
  and fixes,'' \emph{arXiv preprint arXiv:1503.06410}, 2015.

\end{thebibliography}

\end{document}